# Large Magnetoelectric Response in $Sr_2IrO_4$/$SrTiO_3$ superlattices with non-equivalent interfaces

Xin Liu[1], Peipei Lu[2,3], Mei Wu[4,5], Yuanwei Sun[4,5], Jingdi Lu[1], Jing Wang[6], Dayu Yan[2], Youguo Shi[2], Zhiping Yin[1], Nian Xiang Sun[7], Peng Gao[4,5,8,9], Young Sun[2,3], Fa Wang[4,8], Ce-Wen Nan[6] and Jinxing Zhang[1*]

[1]Department of Physics, Beijing Normal University, 100875 Beijing, China

[2]Beijing National Laboratory for Condensed Matter Physics, Institute of Physics, Chinese Academy of Sciences, Beijing 100190, China

[3]School of Physical Science, University of Chinese Academy of Sciences, Beijing 100190, China

[4]International Center for Quantum Materials, Peking University, Beijing 100871, China

[5]Electron Microscopy Laboratory, School of Physics, Peking University, Beijing 100871, China

[6]School of Materials Science and Engineering, Tsinghua University, Beijing 100084, China

[7]Department of Electrical and Computer Engineering, Northeastern University, Boston, Massachusetts 02115, USA

[8]Collaborative Innovation Centre of Quantum Matter, Beijing 100871, China

[9]Beijing Key Laboratory of Quantum Devices, Beijing 100871, China

E-mail: jxzhang@bnu.edu.cn

Large magnetoelectric response in thin films is highly desired for high-throughput and high-density microelectronic applications. However, the $d^0$ rule in single-phase compounds usually results in a weak interaction between ferroelectric and magnetic orders; the magnetoelectric coupling via elastic resonance in composites restricts their thin-film integration in broadband. Here, we effectuate a concurrence of ferroelectric-like and antiferromagnetic phase transitions in $Sr_2IrO_4$/$SrTiO_3$ superlattices by artificial design periodically non-equivalent interfaces, where a maximum magnetoelectric coefficient of ~980 mV cm$^{-1}$ Oe$^{-1}$ can be measured. Evidenced by synchrotron X-ray absorption and electron energy loss spectroscopies, a lopsided electron occupation occurs at the interfacial Ti ions. From perturbative calculations and numerical results, a strong coupling of antiferromagnetism and asymmetric electron occupation mediated by spin-orbit interaction leads to a large bulk magnetoelectric response. This atomic tailoring of the quantum order parameters in $3d$ and $5d$ oxides provides an alternative pathway towards strong magnetoelectric effects with thin-film integrations.

Magnetoelectric (ME) effects, magnetic-field-mediated polarization and/or electric-field-controllable magnetism, have attracted enormous attentions owing to the growing demands of the electronic/spintronic industry[1,2]. The direct ME coupling in single-phase compounds and composite materials can exhibit promising and unique applications in pico-Tesla magnetic sensor, high-efficient transducer and high-throughput electromagnetic communications, etc[3,4]. Single-phase ME materials with various physical origins have been intensely studied over the past few decades[5], such as lone pair ferroelectricity (BiFeO$_3$)[6] and geometric ferroelectricity (YMnO$_3$)[7], etc. However, the $d^0$ rule usually inhibits the coexistence of strong ferroelectricity (favored by empty $d$ orbitals) and ferromagnetism (favored by partially filled $d$ orbitals), which may lead to a small ME response in these multiferroics[8]. Subsequently, spin current[9], exchange striction[10] and $p$-$d$ hybridization[11] models, have been discovered to provide ME coupling under external bias in type-II single-crystal phase multiferroics[12,13], where weak ferroelectricity is induced by spin orders[13,14]. On the other hand, composite materials can exhibit a large ME response at their mechanical resonance frequency. Although sophic theoretical and experimental works have designed and fabricated a plenty of ME composites with excellent performances[15,16], magneto-electro-elastic-driven ME coupling restricts their applications in broadband and integrated thin-film devices[17]. Therefore, there is a strong impetus to explore new ME materials by tailoring the controllable quantum interactions among the spin, orbit, charge and lattice degrees of freedom[18]. Oxide superlattices offer a fertile ground for ferroic structures and functionalities, such as synthetic antiferromagnets[19], enhanced polarization[20], polar vortices[21] and multiferroics[22], which may be also used as an alternative strategy to design strong ME coupling at the atomic scale.

In recent years, spin-orbit coupling (SOC, $\lambda$) in 3$d$ oxides ($\lambda$, ~0.02 eV) has become an effective control parameter to produce polarization[13,23]. Therefore, 5$d$ oxides ($\lambda$, ~0.4 eV) with a variety of emergent quantum and topological states[24,25] may be potential candidates for strong and controllable coupling between magnetism and polarization[26].

Tetragonal $Sr_2IrO_4$ (SIO) with an ~11° staggered $IrO_6$ octahedral rotation about the $c$ axis[27] is one of the Ruddlesden-Popper series $Sr_{n+1}Ir_nO_{3n+1}$ (n=1), which attracts tremendous attentions due to its SOC-induced antiferromagnetic Mott insulator[28,29] and possible high-temperature superconductivity[30]. With a strong SOC and an antiferromagnetic ordering in the $IrO_2$ plane ($T_N$~240 K), SIO provides us a matrix for realizing a possible large ME response. In ME solid-states, the upper bound of the coupling coefficient can be expressed as[31]:

$$\alpha_{ij}^2 \leq \varepsilon_0\mu_0\varepsilon_{ii}\mu_{jj},$$

Where $\alpha_{ij}$ is the ME coefficient (dE/dH)[2,12], $\varepsilon_0$, $\mu_0$, $\varepsilon_{ii}$ and $\mu_{jj}$ are the permittivity of free space, permeability of free space, relative permittivity and relative permeability, respectively. Therefore, quantum paraelectric $SrTiO_3$ (STO) with a high dielectric constant and controllable ferroic properties is a promising option[32,33] to construct building blocks with a large ME response.

Here, SIO/STO superlattices with a non-equivalent interface are artificially designed using laser molecular beam epitaxy (L-MBE) assisted by reflection high-energy electron diffraction (RHEED). Synchrotron X-ray absorption spectroscopy (XAS) and electron energy loss spectroscopy (EELS) demonstrate that an asymmetric electronic structure generates electric dipoles; Temperature-dependent magnetic moment, dielectric constant, piezoresponse force microscopy (PFM) and pyroelectric current show a concurrence of antiferromagnetic and ferroelectric-like phase transitions where the maximum ME coefficient is obtained. Perturbative calculations and numerical results uncover the physical origin: the magnetic field provides spin striction along the $z$-direction in SIO to generate an effective $z$-component Zeeman field, which effectively couples with the electric dipoles across the interface due to its strong SOC. Thus the out-of-plane magnetic-field-controlled electron occupation on Ti creates a large ME response, which is absent when the magnetic field is applied along the $xy$-plane.

$(SIO)_m/(STO)_n$ ($I_m/T_n$) superlattices were constructed on STO (001) single-crystal

substrates (m and n denote the stacking sequence of SIO and STO respectively) by L-MBE with RHEED, where a layer-by-layer growth at the atomic-scale can be achieved. Synchrotron $\theta/2\theta$ X-ray diffractogram (XRD) confirms the periodicity and high-quality epitaxy as shown in Fig. 1a. A 45° rotation about the $c$ axis of the SIO layers on STO (001) substrate is verified, as illustrated in Fig. 1b. A tensile strain of ~0.4% in the SIO layers and the strain-free nature of the STO layers are confirmed by the RHEED patterns (Fig. 1c,d). Figure 1e shows the atomic structure of the $I_3/T_6$ superlattices. By alternatively stacking SrO-IrO$_2$-SrO and TiO$_2$-SrO building blocks, a non-equivalent interface is established in the lattice. As seen in Fig. 1e, only one SrO layer exists within the bottom interface, whereas two SrO layers separate TiO$_2$ and IrO$_2$ at the top interface. This unique structure breaks the space-inversion symmetry, which may be a prerequisite for the formation of an electric dipole[34].

To explore the electronic structure of these asymmetric interfaces, XAS was charactered. Figure 2a shows the XAS of Ti $L_{2,3}$ for $I_3/T_6$, $I_6/T_{12}$ superlattices and STO thin films. Energy difference between the $e_g$ and $t_{2g}$ peaks of the Ti $L_3$ (black) and Ti $L_2$ (red) absorption edges is plotted in Fig. 2b. We can observe that the concetration of $Ti^{3+}$ was dramatically enhanced in the superlattices compared to the one of STO thin film, indicating a strong interface-induced electron occupation at Ti ions[35,36]. EELS further reveals that the lopsided electron occupation occurs at the bottom interface. Hence an electric dipole due to the asymmetric electronic structure forms along the $z$-direction. A temperature-dependent dielectric behavior was respectively measured in these superlattices, SIO and STO thin films. As shown in Fig. 2c, a strain-free STO thin film shows a quantum paraelectric behavior, and the SIO thin film shows no dielectric anomaly. Distinctively, the $I_3/T_6$ superlattice demonstrates a ferroelectric-like phase transition with a broad peak at ~80 K ($T_c$). Low-temperature PFM images and hysteresis at 3.7 K further show that ferroelectric domains are switchable by an external electric field (phase in Fig. 2d, amplitude in Fig. 2e and 180° switching in Fig. 2f). Temperature-dependent PFM images also demonstrate the ferroelectric-like phase transition between 70-90 K, in good accordance with the dielectric measurements. SIO in the superlattices shows an antiferromagnetic behavior, therefore, the coupling between spins and dipoles at the interface may exist in this superlattice.

The standard pyroelectric current measurement[37,38] was carried out to study the direct ME response, the thermo-remanent polarization along the $z$ direction was obtained at various DC magnetic field. The spontaneous polarization emerges, increases rapidly and then saturates as shown in Fig. 3a. While applying the external DC magnetic field along the $z$ direction, the polarization in this superlattice is suppressed near the phase transition, which directly shows the ME coupling effect. To further quantify the ME response, the ME coefficient was measured by a dynamic technique. There is an anisotropic ME response when the magnetic field is applied along and perpendicular to the film plane. When an alternating magnetic field is applied at 10 kHz, the out-of-plane (black curve in Fig. 3b left top) maximum ME coefficient of ~400 mV cm$^{-1}$ Oe$^{-1}$ is obtained in the $I_3/T_6$ superlattice, which is close to the ferroelectric-like (Fig. 3b left middle) and antiferromagnetic phase transitions (Fig. 3b left bottom). The concurrence of ferroelectric-like and antiferromagnetic phase transitions further implies a strong coupling between electric dipoles and spins. In contrast, no direct ME response was observed in STO thin film without ferroelectric and antiferromagnetic phase transitions. In addition, the same dielectric and ME measurements were carried out on the $(Sr_2IrO_4)_3/(LaAlO_3)_6$ ($I_3/A_6$) superlattice as a comparison. No direct ME response was observed in the $I_3/A_6$ superlattice without a ferroelectric-like phase transition. For the $I_6/T_{12}$ superlattice, the ferroelectric-like phase transition is evanescent and away from the appearance of antiferromagnetism[27], leading to a dramatic decrease in the maximum ME coefficient (~110 mV cm$^{-1}$ Oe$^{-1}$ at 10 kHz). When the stacking sequence of SIO and STO further increases (e.g. $I_{12}/T_{24}$), the magnetic and electric orders start to decouple with a small ME coefficient. Frequency-dependent ME coefficient of the $I_3/T_6$ superlattice is shown in Fig. 3c. With increasing frequency of the alternating magnetic field, the ME coefficient gradually increases and saturates (~980 mV cm$^{-1}$ Oe$^{-1}$ at 100 kHz), which is likely due to the dipole relaxation as the frequency-dependent dielectric constant shown (Fig. 3c) according to the capacitance formula[39].

Perturbative calculations and numerical results reveal the microscopic origin of the ME coupling in Fig. 4. The orbital angular momentum of the *d*-electrons is not completely quenched in the Ir atoms, due to the strong SOC compared to the crystal fields breaking cubic symmetry[40]. Our SIO thin film and superlattices show a weak canted magnetic moment within the $IrO_2$ plane, the effective Zeeman field ***B*** on Ir is mainly from the exchange coupling $J \sim 0.1$ eV[41] between the ordered magnetic moments. The Zeeman field along the *z*-direction couples to the *z*-component of spin, and couples to the $|d_{zx}> \pm i|d_{yz}> \propto |L=2, L_z=\pm 1>$ orbitals via the strong SOC[29,40]. Our calculation indicates that occupation of these Ir orbitals will induce occupations of both the Ti $d_{yz}$ and $d_{zx}$ orbitals across the Ir-O-Ti interface as seen in Fig. 4a. However, if the Zeeman field lies along the *x*-direction, it couples to the *x*-component of spin, and couples to the $|d_{xy}> \pm i|d_{zx}> \propto |L=2, L_x=\pm 1>$ orbitals, leading to negligible occupations of only the Ti $d_{zx}$ orbital through the O *p*-orbitals as seen in Fig. 4a. Therefore, the ME response for the Zeeman field in the *xy*-direction is much weaker than that in the *z*-direction field, where an angle-dependent occupations of the Ti orbital across the interface are theoretically obtained in Fig. 4b. This microscopic origin of the ME response is schematically shown in Fig. 4c, where an external magnetic field along the *z*-direction can produce a large effective *z*-component of Zeeman field, which deforms the occupied states on the Ir, thus reducing the electron occupation on Ti to induce a significant ME response via the SOC. Nevertheless, the moments will remain in the *xy*-plane under an in-plane magnetic field, so no significant ME response will be produced (Fig. 4d). In practice, the spin striction is the most sensitive at the antiferromagnetic phase transition, where the maximum ME response occurs as shown in Fig. 3a and 3b.

In summary, a large bulk ME coefficient of ~980 mV cm$^{-1}$ Oe$^{-1}$ was obtained by the atomic tailoring of the periodical non-equivalent interfaces between SIO with a strong

SOC and quantum paraelectric STO, where a concurrence of ferroelectric-like and antiferromagnetic phase transitions occurs. This ME response mediated by interfacial SOC indicates that these artificially designed superlattices are promising candidates to be applied in integrated devices without an assistance of DC magnetic bias. Furthermore, atomic design of the superstructures in layered- (e.g. $A_2BO_4$, $A_3B_2O_7$) and traditional-perovskite ($ABO_3$) oxides also naturally provides us a new avenue towards realizing non-centrosymmetric systems with strong electron correlation for exploring emergent quantum phenomena such as unconventional superconductivity and topological order states, etc.

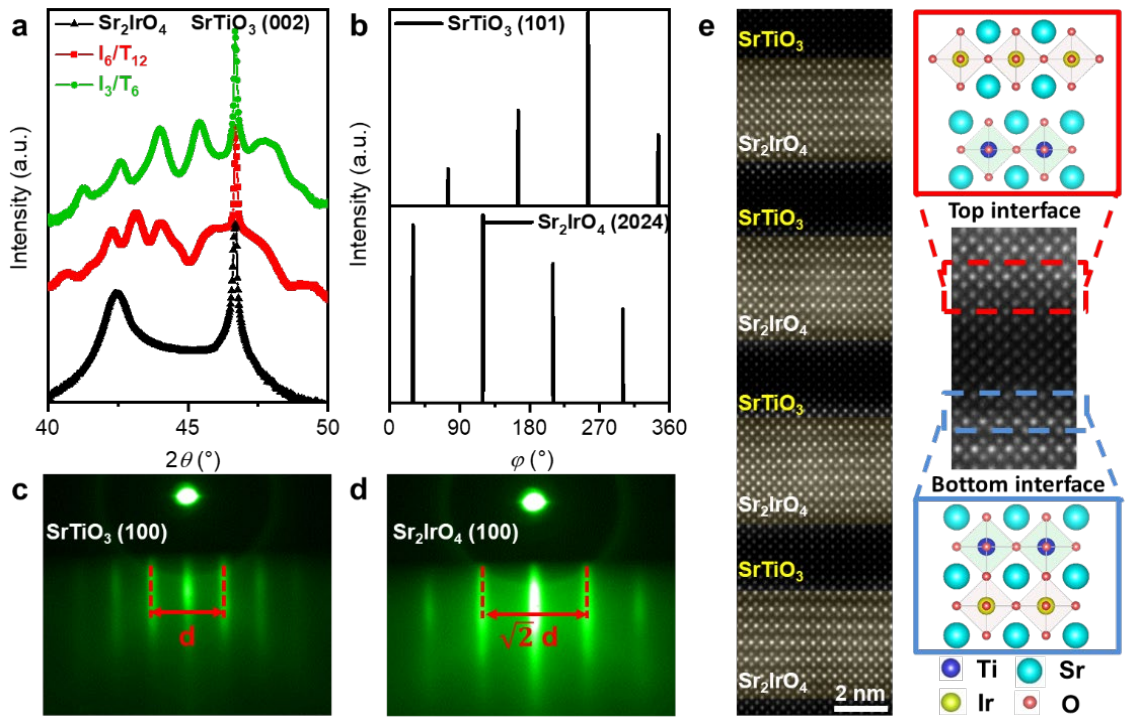

**Figure 1 | Atomic design of (SIO)$_m$/(STO)$_n$ superlattices with non-equivalent interfaces. a,** $\theta$-$2\theta$ X-ray diffractogram for superlattices and SIO thin film on STO (100) substrate, confirming the periodicity and high crystallinity. **b,** X-ray diffraction azimuthal $\varphi$ scans of SIO (2024) and STO (100). There is a 45° rotation of the SIO layers about the $c$ axis on STO. **c, d,** RHEED patterns for STO (100) and SIO (100), respectively, verifying the 45° rotation and high quality of the superlattice with a tensile strain of ~0.4% for SIO. **e,** Cross-sectional high-angle annular dark-field of scanning transmission electron microscopy (HAADF-STEM) image and schematic structure of the I$_3$/T$_6$ superlattice. An asymmetric interface is artificially constructed.

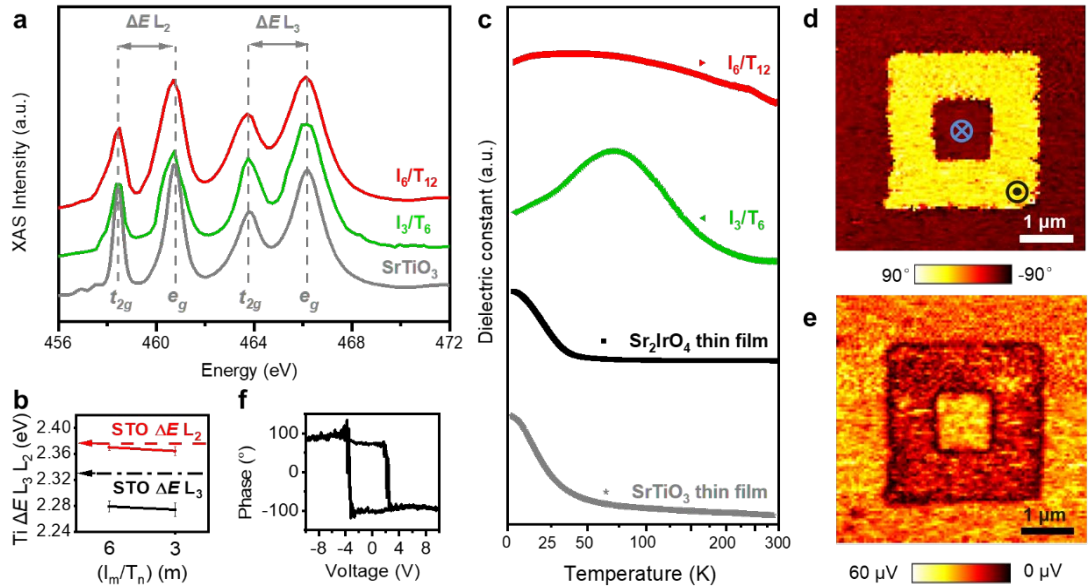

**Figure 2 | Interfacial electron occupation and ferroelectric-like transition. a,** Ti $L_{2,3}$ X-ray isotropic absorption spectra for superlattices and STO thin film. **b,** The energy difference between the $e_g$ and $t_{2g}$ peaks of the Ti $L_3$ (black) and Ti $L_2$ (red) absorption edges, where the black and red dash-dotted arrows represent the energy difference in the STO thin film. **c,** Temperature-dependent dielectric constant for superlattices, SIO and STO thin films along the out-of-plane at 10 kHz. There is no dielectric anomaly for the strain-free STO thin film indicating a quantum paraelectricity. The SIO thin film shows no dielectric anomaly, however, a broader peak at 80 K was measured for the $I_3/T_6$ superlattice implying a relaxor-like behavior. **d, e,** Out-of-plane PFM image of the $I_3/T_6$ superlattice at 3.7 K, the relative dark and bright contrasts in phase **(d)** and amplitude **(e)** indicate upward and downward ferroelectric domains. **f,** Piezoresponse phase hysteresis of the $I_3/T_6$ superlattice confirms the switchable domains.

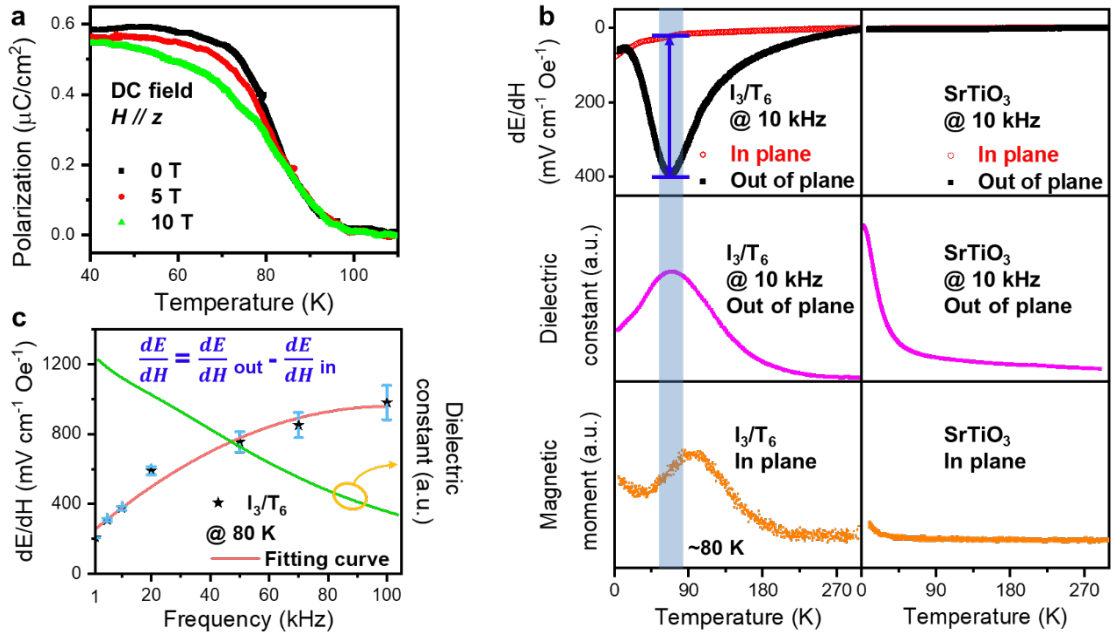

**Figure 3 | Direct ME effect and concurrence of ferroelectric-like and antiferromagnetic transitions. a,** Temperature dependence of electric polarization under external DC magnetic field along the *z* axis. The polarization was obtained by integration of pyroelectric current. **b,** Direct measurements of the temperature-dependent coefficient (top) for the $I_3/T_6$ superlattice and STO thin film at 10 kHz. The out-of-plane dielectric (middle) and in-plane moment (bottom) behaviors as a function of temperature are plotted for comparison. There is an anisotropic ME response when a magnetic field is applied along and perpendicular to the film plane. The maximum ME coefficient is close to the ferroelectric-like transition and the formation of antiferromagnetism implying strong coupling between electric dipoles and spins. In contrast, no direct ME response was observed in the STO thin film without ferroelectric-like and antiferromagnetic phase transitions. **c,** ME coefficient of the $I_3/T_6$ superlattice at 1, 5, 10, 20, 50, 70 and 100 kHz respectively, where an AC magnetic field is applied along the out-of-plane. The coefficient increases and saturates gradually as the fitting curve shown. A maximum coefficient of ~980 mV cm$^{-1}$ Oe$^{-1}$ was obtained when applying a alternating magnetic field (~2.5 Oe) at 100 kHz. Frequency-dependent dielectric constant (green line) shows a inversely proportional tendency with the ME coefficient.

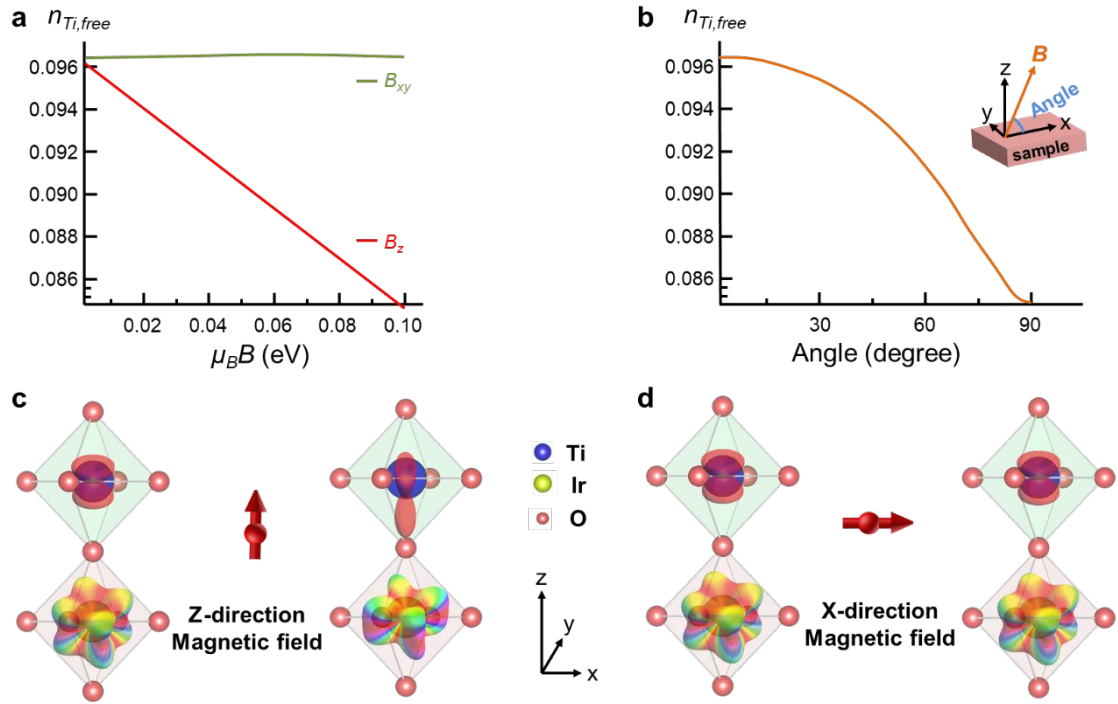

**Figure 4 | Mechanism of magnetoelectric coupling. a, b,** Anistropy of ME coupling for the free fermion model of numerical results. **(a)** Occupation of Ti *d*-orbitals $n_{Ti}$ versus the angle between external magnetic field ***B*** and the *x*-direction as the inset shown. **(b)** Occupation of Ti *d*-orbitals $n_{Ti}$ versus the effective Zeeman field on the Ir along the *z*-direction and in the *xy*-plane. **c, d,** Schematic picture of the electronic states on the Ir and Ti atoms under external magnetic fields along the *z*-direction and *x*-direction. The bottom row indicates occupied $J_{\text{eff}}= 1/2$ states on the Ir atoms, the top row indicates occupied spinful combinations of $t_{2g}$ orbitals on the Ti atoms. External magnetic fields along the *z*-direction induces strong Zeeman fields of *z*-component that deforms the occupied states on the Ir, thus reducing the electron occupation on Ti via SOC as shown in **(c)**. External magnetic fields in the *xy*-plane have no effect on ME coupling as shown in **(d)**.

# Acknowledgements

The work at Beijing Normal University is supported by the National Key Research and Development Program of China through Contract 2016YFA0302300. J.Z. also acknowledges the support from "The Fundamental Research Funds for the Central Universities" under Contracts 2017EYT26 and 2017STUD25. Y.S. acknowledges the support from the National Natural Science Foundation of China under Grant No. 51725104. P.G. also acknowledges the support from the National Natural Science Foundation of China (51672007). We gratefully acknowledge Electron Microscopy Laboratory in Peking University for the use of Cs corrected electron microscope. F.W. acknowledges support from The National Key Research and Development Program of China (Grand No. 2017YFA0302904).

## Author contributions

J.Z. and X.L. designed the experiments and prepared the manuscript. X.L. carried out the growth of all the samples. The polycrystalline target was synthesized by D.Y. and Y.S.. Synchrotron X-ray diffraction, synchrotron X-ray absorption spectroscopy and temperature/frequency-dependent dielectric behaviors were performed by X.L.. J.L. and X.L. carried out the temperature-dependent PFM and AFM measurements. Y.S. recorded the STEM images and M.W. did the EELS measurement under the direction of P.G.. X.L. and J.W. carried out the SQUID measurement. Pyroelectric current and ME coefficient measurements were performed by X.L. and P.L. Perturbative calculations and Numerical results of theoretical model were carried out by F.W.. C.-W.N., Z.Y., N.X.S., P.G., and F.W. were involved in the discussion and revision of the manuscript. All authors were involved in the analysis of the experimental and theoretical results.